\documentclass{article}
\usepackage{spconf,amsmath,graphicx}

\usepackage{enumitem}
\usepackage{booktabs}
\usepackage{amsmath}
\usepackage{amssymb}
 
\setlist{nosep, leftmargin=14pt}

\title{Imaging the Topology of Dynamic Brain Connectivity}

\name{Peilin He$^{1}$, Tananun Songdechakraiwut$^{2}$}
\address{$^{1}$ Division of Natural and Applied Sciences, Duke Kunshan University \\ \quad $^{2}$ Department of Computer Science, Duke University}

\begin{document}

\maketitle

\begin{abstract}
Functional brain connectivity changes dynamically over time, making its representation challenging for learning on non-Euclidean data. We present a framework that encodes dynamic functional connectivity as an image representation of evolving network topology. Persistent graph homology summarizes global organization across scales, yielding Wasserstein distance-preserving embeddings stable under resolution changes. Stacking these embeddings forms a topological image that captures temporal reconfiguration of brain networks. This design enables convolutional architectures and transfer learning from pretrained foundational models to operate effectively under limited and imbalanced data. Applied to early Alzheimer's detection, the approach achieves clinically meaningful accuracy, establishing a principled foundation for imaging dynamic brain topology.
\end{abstract}


\section{Introduction}

Learning effective representations of functional brain connectivity remains a central challenge in computational neuroimaging. Functional networks exhibit complex, time-varying organization defined on non-Euclidean domains, which poses a representational bottleneck for conventional deep models optimized for grid-structured data. Existing correlation- and graph-based features capture local interactions but underrepresent higher-order structure and temporal reconfiguration, limiting generalization under the data scarcity and class imbalance typical of clinical neuroimaging.

We introduce a framework that transforms time-resolved functional connectivity graphs into a compact image-based representation capturing the dynamically changing topology of brain networks. Using persistent graph homology~\cite{songdechakraiwut2023topological}, the method summarizes global network organization across connectivity scales, producing Wasserstein distance-preserving embeddings that remain stable under varying resolutions~\cite{skraba2020wasserstein}. Stacking these embeddings sequentially forms a two-dimensional topological image that captures the evolution of network topology over time.

This transformation bridges non-Euclidean brain network analysis with modern computer vision, enabling convolutional architectures and transfer learning from pretrained foundational models~\cite{Mienye2025_DeepCNNMedImage} to capture functional dynamics effectively under limited and imbalanced data conditions. Applied to early Alzheimer's disease detection on OASIS-3~\cite{LaMontagne2019_OASIS3}, the framework achieves clinically meaningful prediction accuracy on dementia ratings~\cite{muir2024minimal}, demonstrating sensitivity to subtle functional disruptions preceding structural decline. The proposed approach establishes a principled foundation for imaging and analyzing the topology of dynamic brain connectivity.

\section{Topological Image Representation}

Functional magnetic resonance imaging (fMRI) measures spontaneous fluctuations in the blood-oxygen-level-dependent (BOLD) signal, which reflect hemodynamic responses coupled to neural activity across the brain. The brain is parcellated into $|V|$ regions of interest (ROIs) using a predefined atlas, and the mean BOLD signal within each ROI is extracted to form regional time series. To examine how functional interactions evolve over time, we employ a \emph{sliding-window approach} that partitions the full BOLD time series into overlapping temporal windows. Within each window, pairwise functional correlations between ROIs are calculated, yielding a sequence of time-resolved \emph{functional connectivity matrices} that describe the evolving organization of brain networks.

Each connectivity matrix is represented as an undirected, weighted graph $G = (V, W)$, where $V$ denotes brain regions and $W = [w_{ij}] \in \mathbb{R}^{|V| \times |V|}$ is a symmetric weight matrix encoding the strength of pairwise functional connections. To extract global and interpretable structure from these graphs, we apply persistent graph homology (PGH)~\cite{songdechakraiwut2023topological}, which quantifies the evolution of topological invariants, connected components (0-homology), and cycles (1-homology), across all edge-weight thresholds using closed-form computation.

For each graph $G$, we construct a family of thresholded subgraphs $G_{\epsilon}$ by retaining edges whose weights exceed a threshold $\epsilon$. Increasing $\epsilon$ generates a \emph{filtration} of nested graphs:
$G_{\epsilon_0} \supseteq G_{\epsilon_1} \supseteq \cdots \supseteq G_{\epsilon_k},$
where $\epsilon_0 \leq \epsilon_1 \leq \cdots \leq \epsilon_k$ are filtration levels. PGH tracks the birth and death of topological features across this filtration. Each feature that appears at $b_l$ and disappears at $d_l$ corresponds to a point $(b_l, d_l)$ in a persistence diagram~\cite{edelsbrunner2022computational}.
Because the number of connected components ($\beta_0$) increases and the number of independent cycles ($\beta_1$) decreases monotonically with $\epsilon$ ~\cite{songdechakraiwut2023topological}, the topology of $G$ can be summarized by the \emph{birth times} of connected components and the \emph{death times} of cycles:
$B(G) = \{ b_l \}_{l=1}^{|V|-1}, 
D(G) = \{ d_l \}_{l=1}^{1 + |V|(|V|-3)/2}.$

To obtain a smooth and resolution-flexible representation, we employ \emph{inverse transform sampling} to interpolate the empirical distributions of $B(G)$ and $D(G)$.  
For the birth set, the empirical distribution is
$f_{G,B}(x) = \frac{1}{|B(G)|} \sum_{b \in B(G)} \delta(x-b),$
where $\delta(x-b)$ denotes a Dirac delta function centered at $b$.
The corresponding cumulative function is
$F_{G,B}(x) = \frac{1}{|B(G)|} \sum_{b \in B(G)} \mathbf{1}_{b \leq x},$
and its pseudo-inverse is defined as
$F_{G,B}^{-1}(z) = \inf\{ b \in \mathbb{R} \mid F_{G,B}(b) \geq z \}.$
Sampling $F_{G,B}^{-1}$ at $m$ uniform quantiles yields a fixed-length, continuously resampled embedding:
\[
\mathbf{v}_{B} = 
\big( F^{-1}_{G,B}(\tfrac{1}{m}), \ldots, F^{-1}_{G,B}(\tfrac{m}{m}) \big)^\top,
\]
and analogously for the death set:
\[
\mathbf{v}_{D} = 
\big( F^{-1}_{G,D}(\tfrac{1}{n}), \ldots, F^{-1}_{G,D}(\tfrac{n}{n}) \big)^\top.
\]
This inverse-transform-based interpolation enables the topological representation to be \emph{resampled at arbitrary resolutions} (by varying $m$ or $n$) while \emph{preserving} Wasserstein distances between persistence diagrams~\cite{songdechakraiwut2023wasserstein}, ensuring topological stability under changes in sampling resolution~\cite{skraba2020wasserstein}.

\begin{figure}[t]
    \centering
    \includegraphics[width=0.5\textwidth]{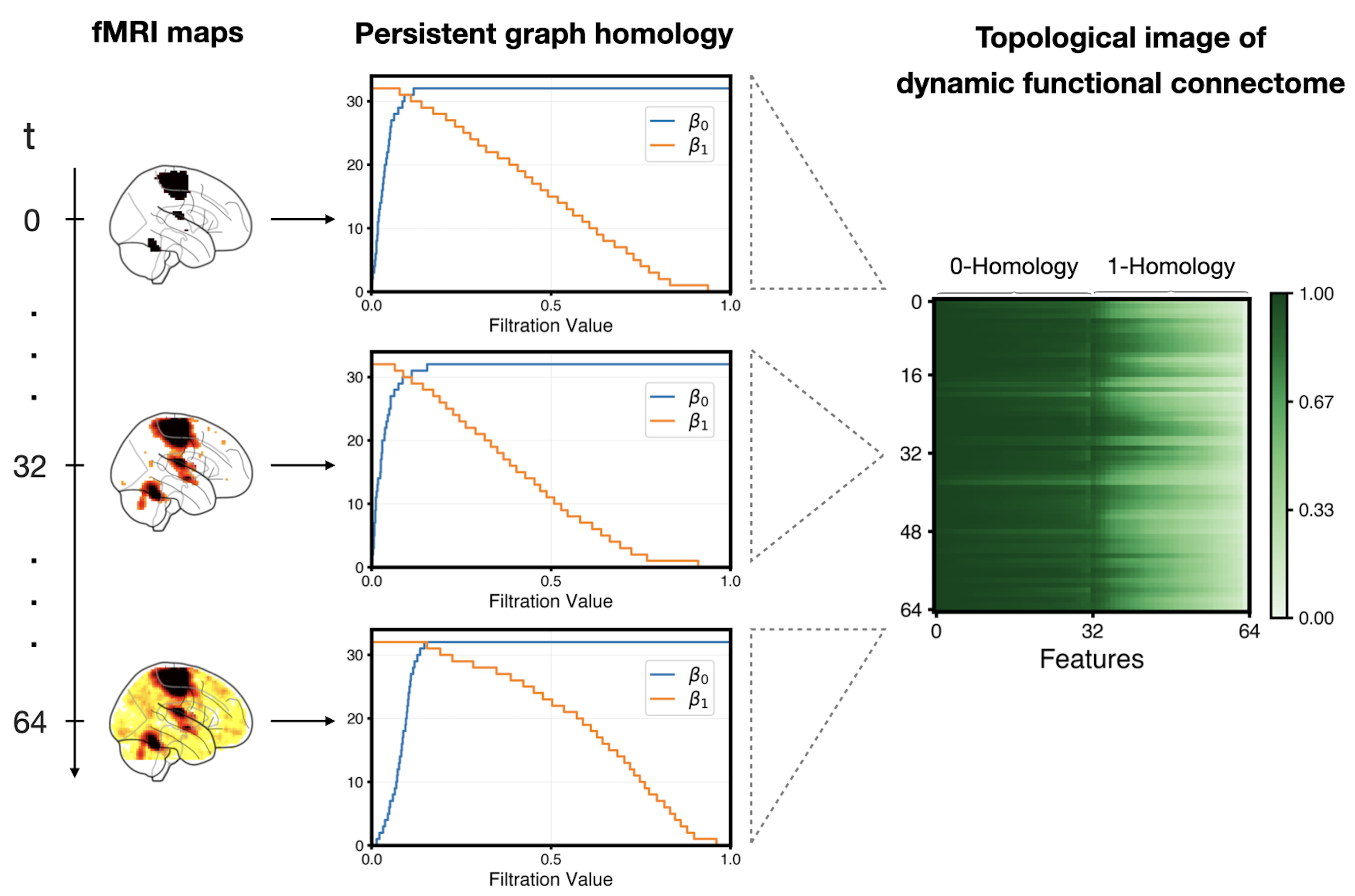}
    \caption{%
    Topological representation of a dynamic functional connectome.
    At each time point $t$ (left), fMRI activation maps are converted into
    functional connectivity graphs and summarized via persistent graph homology
    (middle), yielding Betti curves for $0$- and $1$-dimensional homology
    ($\beta_0$, $\beta_1$) over the filtration values. Stacking the
    resulting topological summaries across time produces a 2D topological
    image (right), where rows correspond to time windows and columns encode
    features from $0$- and $1$-homology.
    }
    \label{fig:topo-dynamic-connectome}
\end{figure}

We then concatenate the two embeddings to form a unified topological feature vector:
\[
F(G) = [\mathbf{v}_{B} \; || \; \mathbf{v}_{D}],
\]
which jointly encodes the 0- and 1-dimensional homological features of the graph.  
For a sequence of sliding windows $\{G_t\}_{t=1}^{T}$, the corresponding feature vectors are concatenated over time:
\[
\mathcal{I} = [F(G_1), F(G_2), \ldots, F(G_T)] \in \mathbb{R}^{(m+n) \times T}.
\]
The resulting \emph{image-like representation} $\mathcal{I}$ captures the temporal evolution of the brain's topological connectivity in a compact and interpretable form.
Each column of $\mathcal{I}$ represents the topological feature vector $F(G_t)$ from a specific time window, while each row traces the temporal evolution of a sampled topological scale. 
Viewed as a two-dimensional map, the horizontal axis encodes time and the vertical axis encodes persistence scale. 
Smooth regions indicate stable network topology, whereas abrupt or localized intensity changes mark dynamic reconfigurations of persistent connected components and cycles. An overview of the proposed pipeline is shown in Fig.~\ref{fig:topo-dynamic-connectome}.

\section{Application to Early Detection of Alzheimer's Disease}

\subsection{Datasets}

\begin{table}[t]
\centering
\small
\begin{tabular}{ll}
\toprule
\textbf{Item} & \textbf{Setting} \\
\midrule
Learning rates & Backbone $1\times 10^{-4}$; Head $2\times 10^{-4}$ \\
Gradient clipping & $\ell_2$-norm $=5.0$ \\
Preprocessing & \texttt{.npy} $\to$ $[0,255]$ \texttt{uint8} \\
Augmentation & Rotation $\le 5^\circ$, translation $\le 5\%$ \\
\bottomrule
\end{tabular}
\caption{Training hyperparameters.}
\label{tab:train_hparams}
\end{table}

We evaluate our representation framework using the OASIS-3 dataset~\cite{LaMontagne2019_OASIS3}, which provides longitudinal, multimodal neuroimaging and detailed clinical assessments. The cohort comprises 839 participants with resting-state fMRI and corresponding diagnostic labels: 772 with normal cognition (N), 34 with mild cognitive impairment (MCI), and 33 with cognitive impairment not meeting MCI criteria (IMP). All fMRI data were preprocessed using fMRIPrep, following a standardized and reproducible pipeline consistent with our previous work~\cite{yi2025topological}. Subjects with dementia were excluded to focus on early-stage cognitive decline relevant to Alzheimer's disease detection. Among the 34 MCI cases, one subject with an extreme Clinical Dementia Rating Sum of Boxes (CDR-SB) score (9.0) was treated as an outlier and excluded. To capture this early impairment spectrum, the MCI and IMP groups were merged into a single category (IMP+MCI) and contrasted with the normal cohort.

\subsection{Model configurations}

\begin{table*}[t]
\centering
\footnotesize
\begin{tabular}{lccc}
\toprule
\textbf{Model} & \multicolumn{2}{c}{\textbf{MAE (95\% CIs)}} & $\mathbf{MAE}_{\text{entropy}}$ \\
\cmidrule(lr){2-3}
 & \textbf{IMP+MCI} & \textbf{N} & \\
\midrule
EfficientNet-B0           & 0.6513 (0.4471, 0.8783) & 0.5806 (0.5729, 0.5878) & 0.6316 \\
ResNet-18                 & 0.7064 (0.5003, 0.9472) & 0.4556 (0.4484, 0.4624) & 0.6367 \\
DenseNet-121              & 0.6517 (0.4820, 0.8460) & 0.3848 (0.3777, 0.3915) & 0.5775 \\
ResNet-34                 & 0.6937 (0.4866, 0.9300) & 0.4770 (0.4697, 0.4840) & 0.6333 \\
ResNet-50                 & 0.6887 (0.5247, 0.8707) & 0.3479 (0.3427, 0.3529) & 0.5939 \\
\bottomrule
\end{tabular}
\caption{\textbf{CDR--SB regression on the test set (33\% frozen fraction).} Mean absolute error (MAE) with 95\% bootstrap CIs by diagnosis, together with an entropy-weighted combined MAE. }
\label{tab:mae}
\end{table*}

\paragraph*{Outlier-weighted mean squared error loss.}
To improve sensitivity to underrepresented but clinically important large-target cases (i.e., the positive diagnostic cases), we train with an outlier-weighted mean squared error (OW-MSE). Let $(\hat y_i, y_i)$ denote the prediction and ground-truth for sample $i$. We first compute the $\tau$-quantile of the targets on the training set, $q_\tau = \mathrm{Quantile}_\tau\big(\{y_j\}_{j \in \text{train}}\big)$, and assign a piecewise-constant weight $w_i = w$ if $y_i > q_\tau$ and $w_i = 1$ otherwise. The loss is then defined as
$\mathcal{L}_{\text{OW-MSE}} = \tfrac{1}{N} \sum_{i=1}^{N} w_i\,(\hat y_i - y_i)^2$. In our study, we use $\tau = 0.90$ and $w = 3.0$.

\paragraph*{Architecture and optimization.}
 
We evaluate our method using five Convolutional Neural Networks (CNNs): DenseNet-121, EfficientNet-B0, ResNet-18, ResNet-34, and ResNet-50~\cite{Mienye2025_DeepCNNMedImage}. For ResNet-18/34/50, we use a small-stem variant (initial $3{\times}3$ convolution with stride $1$ and no first max-pooling layer) to better preserve local spatial detail on our low-resolution inputs. All models are trained on single-channel inputs for 60 epochs using AdamW. We use a 5-epoch linear warm-up for the learning rate, followed by cosine decay with a minimum of $1\times 10^{-6}$; the weight decay is set to $3\times 10^{-4}$. We adopt \emph{partial fine-tuning}: early backbone stages are frozen (approximately $x\%$ of parameters), and only the final stage and task head are updated (see Section~\ref{sec:ablation_frozen} for ablations on the frozen depth of the backbone). To mitigate class imbalance, we use a rebalanced batch sampler on the training split, enforcing a per-batch IMP{+}MCI fraction of 0.65 via independent draws with replacement from the IMP{+}MCI and N pools. The remaining hyperparameters (e.g., separate backbone/head learning rates, gradient clipping, preprocessing, and augmentation) are summarized in Table~\ref{tab:train_hparams}.

\subsection{Results and clinical relevance}

Our primary regression target is the CDR-SB, a widely used quantitative index
of cognitive and functional impairment in both research and clinical practice.
We split the dataset by subject (ensuring no overlap across partitions) into
5\% training, 15\% validation, and 80\% test sets (see
Section~\ref{sec:ablation_train} for ablations on training size). Given the limited number of MCI and IMP cases, this design prioritizes subject-level independence and a large held-out test set for robust generalization assessment. For all reported metrics, 95\% confidence intervals (CIs) are estimated via 5{,}000 bootstrap resamples of the test set.

To summarize performance under class imbalance, we introduce an entropy-weighted combined mean absolute error (MAE). Specifically, for $g\in\{\text{N},\text{IMP+MCI}\}$, let $p_g= n_g / (n_N+n_{\text{IMP+MCI}})$, $h_g=-p_g\log p_g$, and $w_g = h_g / (h_N + h_{\text{IMP+MCI}})$.
The combined score is then
$\mathrm{MAE}_{\text{entropy}} = w_N\,\mathrm{MAE}_N + w_{\text{IMP+MCI}}\,\mathrm{MAE}_{\text{IMP+MCI}}$.
Table~\ref{tab:mae} reports MAE with 95\% CIs for each model backbone, evaluated separately on the IMP+MCI and N groups, together with the entropy-weighted combined MAE. Relative to the minimum clinically important difference (MCID) for CDR-SB in mildly impaired populations (1.0--2.0 points~\cite{muir2024minimal}), all IMP+MCI MAEs fall below 1.0 (lie between $0.65$ and $0.71$), and even the upper bounds of their 95\% CIs remain under 1.0 (no greater than $0.95$), suggesting potential clinical utility for early monitoring and disease staging. The N group achieves lower MAEs (between $0.35$ and $0.60$) with markedly narrow CIs (about $\pm 0.01$), consistent with greater precision arising from a larger sample size and lower within-group heterogeneity. Across backbones, performance differences in the IMP+MCI group are less distinct due to overlapping CIs, whereas in the N group, non-overlapping CIs suggest architecture-dependent variations in predictive precision under our evaluation protocol. Aggregated across diagnoses via the entropy-weighted MAE, all models achieve scores between 0.57 and 0.64, remaining below one CDR-SB point on average, suggesting that our
conclusions remain consistent when summarized with this class-imbalance-aware metric.

Early-stage detection of Alzheimer's disease is a long-standing challenge, as clinical symptoms emerge gradually and macrostructural atrophy typically occurs only after earlier network-level dysfunction. By constructing topological image representations of dynamic functional connectivity, our framework quantifies temporally resolved changes in brain network topology that may reflect early synaptic and connectivity alterations preceding structural degeneration. When modeled with convolutional architectures, these representations achieve sub-MCID prediction accuracy on mildly impaired individuals, which is consistent with sensitivity to subtle functional disruptions associated with prodromal Alzheimer's disease. Taken together, these findings suggest that topologically informed functional representations can improve the specificity of learning-based biomarkers and provide a promising pathway toward earlier detection and longitudinal tracking of Alzheimer's progression.

\subsection{Ablation analyses}

In this study, we assess how two design choices affect performance, measured by MAE with 95\% bootstrap CIs: (i) the fraction of the backbone frozen during training and (ii) the proportion of subjects used for training.

\paragraph*{Effect of backbone frozen depth.}
\label{sec:ablation_frozen}

\begin{table}[t]
\centering
\footnotesize
\begin{tabular}{c|cc}
\toprule
\textbf{Frozen fraction \%} & \multicolumn{2}{c}{\textbf{MAE (95\% CIs)}} \\
\cmidrule(lr){2-3}
 & \textbf{IMP+MCI} & \textbf{N} \\
\midrule
0\%   & 0.7193 (0.5120, 0.9732) & 0.4438 (0.4369, 0.4502) \\
17\%  & 0.7086 (0.4914, 0.9500) & 0.4534 (0.4463, 0.4600) \\
33\%  & 0.7064 (0.5003, 0.9472) & 0.4556 (0.4484, 0.4624) \\
50\%  & 0.6614 (0.4584, 0.8922) & 0.5354 (0.5273, 0.5427) \\
67\%  & 0.6545 (0.4545, 0.8971) & 0.5502 (0.5420, 0.5578) \\
90\%  & 0.6463 (0.4507, 0.8734) & 0.5679 (0.5602, 0.5752) \\
100\% & 0.6027 (0.4223, 0.8217) & 0.6929 (0.6847, 0.7002) \\
\bottomrule
\end{tabular}
\caption{Effect of freezing fraction on ResNet-18: MAE with 95\% bootstrap CIs.}
\label{tab:ablation_frozen}
\end{table}

Table~\ref{tab:ablation_frozen} shows that MAE remains below 1.0 across all frozen fractions for both diagnostic groups. As the frozen fraction increases, MAE for IMP+MCI decreases from 0.7193 to 0.6027 with overlapping 95\% CIs, whereas MAE for N increases from 0.4438 to 0.6929 with narrow CIs, illustrating a trade-off between performance on impaired and normal subjects.

\paragraph*{Effect of training set size.}
\label{sec:ablation_train}
\begin{table}[t]
\centering
\small
\begin{tabular}{c|cc}
\toprule
\textbf{Train \%} & \multicolumn{2}{c}{\textbf{MAE (95\% CIs)}} \\
\cmidrule(lr){2-3}
 & \textbf{IMP+MCI} & \textbf{N} \\
\midrule
1\%  & 0.7951 (0.5086, 1.1622) & 0.5718 (0.5625, 0.5806) \\
5\%  & 0.6513 (0.4471, 0.8783) & 0.5806 (0.5729, 0.5878) \\
10\% & 0.7001 (0.4154, 1.0834) & 0.7491 (0.7399, 0.7578) \\
\bottomrule
\end{tabular}
\caption{Effect of training set size on EfficientNet-B0: MAE with 95\% bootstrap CIs.}
\label{tab:ablation_train_size}
\end{table}

Table~\ref{tab:ablation_train_size} shows that, among the three training fractions evaluated, using 5\% of the dataset provides the best overall balance between the N and IMP+MCI cohorts. Increasing the split to 10\% does not yield further gains and instead worsens MAE in both groups, indicating that, with our current training setup, more data does not necessarily improve generalization.

\section{Compliance with ethical standards}
This research study was conducted retrospectively using human subject data made available in open access by OASIS-3 project~\cite{LaMontagne2019_OASIS3}. Ethical approval was not required as confirmed by the license attached with the open access data.

\section{Conflicts of interest}
This work was supported by Duke Science and Technology. The authors declare no competing interests.

\bibliographystyle{IEEEbib}
\bibliography{refs}

\end{document}